# Several Populations of Sunspot Group Numbers – Resolving a Conundrum

## Leif Svalgaard[1*]

[1]W.W. Hansen Experimental Physics Laboratory, Stanford University
Cypress Hall, C3, 466 Via Ortega, Stanford, CA 94305-4085
[*]Corresponding author: Leif Svalgaard (leif@leif.org)

**ABSTRACT**

The long-standing disparity between the sunspot number record and the Hoyt and Schatten (1998, H&S) Group Sunspot Number series was initially resolved by the Clette et al. (2014) revision of the sunspot number and the group number series. The revisions resulted in a flurry of dissenting group number series while the revised sunspot number series was generally accepted. Thus, the disparity persisted and confusion reigned, with the choice of solar activity dataset continuing to be a free parameter. A number of workshops and follow-up collaborative efforts by the community have not yet brought clarity. We review here several lines of evidence that validate the original revisions put forward by Clette et al. (2014) and suggest that the perceived conundrum no longer need to delay acceptance and general use of the revised series. We argue that the solar observations constitute several distinct populations with different properties which explain the various discontinuities in the series. This is supported by several proxies: diurnal variation of the geomagnetic field, geomagnetic signature of the strength of the heliomagnetic field, and variation of radionuclides. The Waldmeier effect shows that the sunspot number scale has not changed over the last 270 years and a mistaken scale factor between observers Wolf and Wolfer explains the disparity beginning in 1882 between the sunspot number and the H&S reconstruction of the group number. Observations with replica of 18[th] century telescopes (with similar optical flaws) validate the early sunspot number scale; while a reconstruction of the group number with monthly resolution (with many more degrees of freedom) validate the size of Solar Cycle 11 given by the revised series that the dissenting series fail to meet. Based on the evidence at hand, we urge the working groups tasked with producing community-vetted and agreed-upon solar activity series to complete their work expeditiously.

## 1. Introduction

At the centenary of Rudolf Wolf's death, Hoyt et al. (1994) asked "Do we have the correct reconstruction of solar activity" and proceeded to answer in the negative by introducing a new reconstruction of solar activity (Hoyt and Schatten (1998; H&S from now on)) as a modern improvement of the Sunspot Number series originally instigated by Wolf (1851) and using in literally thousands of studies of the sun and its effects on the Earth and its environment. Unfortunately, those two series did not match before 1882 AD, resulting in confusion and disagreements, e.g. when used in studies of the solar dynamo or of solar forcing on climate, where the choice of solar activity series now became, essentially, a free parameter. In an attempt to remedy this, a series of Sunspot Number Workshops was initiated with attendance from stake-holders and community-experts (Cliver et al., 2013, 2015,). The hoped-for goal of this effort was a community-vetted time series of sunspot (and group) numbers for use in long-term studies [http://ssnworkshop.wikia.com/wiki/Home].

1.1 Sunspot Number Workshops

The SSN workshops were sponsored by the National Solar Observatory (NSO), the Royal Observatory of Belgium (ROB), and the Air Force Research Laboratory (AFRL). Each workshop was attended by 20-50 participants drawn from both observers and user-communities. A special Topical Issue of the Solar Physics journal with 36 historical, procedural, and research papers resulted from the workshops (Clette et al., 2016). A synthesis of the work was presented to the IAU at its XXIX General Assembly in 2015 (Clette et al., 2014) and is now under the aegis of IAU. However, the discrepancy between the traditional sunspot number and the newer Group Sunspot Number of H&S was not resolved. Nevertheless, the changes to the sunspot number series were less controversial and the World Data Center for Sunspot Index and Long-term Solar Observations (WDC-SILSO) in Brussels could issue a new and updated version of SN: the Sunspot Number version 2 (Clette and Lefèvre, 2016, 2018).

Svalgaard and Schatten (2016) constructed a series of yearly sunspot-group counts, not just by comparisons with other reconstructions and correcting those where they were deemed to be deficient, but by a complete re-assessment from original sources. The resulting solar activity series, now called just the Group Number, GN, generally agreed well with the revised sunspot number series and appeared to reach and sustain for extended intervals of time the same level of activity in each of the last three centuries since 1700 AD and the past several decades did not seem to have been exceptionally active, contrary to what H&S had claimed and what many researchers had been led to accept.

Instead of a hoped-for unified, community-vetted, and agreed upon modern solar activity reference series, a (large) number of dissenting (mainly for the group number) series disagreeing with the adopted version 2 of the sunspot number kept appearing. As Cliver and Herbst (2018) noted "The situation facing the solar community in 2016 was thus scientifically complicated and, on a human level, becoming increasingly contentious. The danger was that the proliferation of new disparate series, if left unaddressed in a systematic fashion, would render the sunspot number meaningless as a measure of solar activity". A recent attempt to reconcile the various series by an International Space Science Institute (ISSI) "International Team 417" (Pesnell et al., 2020) has not brought clarity and no resolution seems in sight. The present article should be seen as a contribution towards restarting that stalled reconciliation effort.

**2. Data and Methods**

The great service performed by H&S was the compilation of a database of all the raw daily observations that was used in constructing the GSN, organized by year and observer in easy-to-read textual format. The database (although in a different format), augmented with observations that have been recovered since the H&S glorious effort, is now curated by Vaquero et al. (2016). The newest version at http://haso.unex.es/haso/index.php/on-line-archive/data/ (v1.21, dated 2020-04-19) with more than a million observations form the base material for the present study.

Already Svalgaard and Schatten (2016) pointed out how remarkable it is that the raw data (that is, simply the yearly average of all observations by all observers) with no normalization at all closely match (coefficient of determination for linear regression $R^2 = 0.97$) the number of groups calculated by dividing the H&S GSN by an appropriate scale factor (14), demonstrating that the elaborate and obscure normalization procedures employed by H&S have almost no effect on the result. "The normalization thus did not introduce, remove, or correct any trends (such as the 'secular increase' from 1700 to the present) or anomalies that were not already in the raw data". The top panel of Figure 1 shows the yearly averages of all observations from the earliest crude telescopes to high-quality modern instruments underpinned by the ever increasing understanding of "the longest running experiment in physics" (Owens, 2013). It is easy to see how the (perhaps politically expedient) notion of steadily increasing solar activity (to become "the largest in 11,000 years", e.g. Solanki et al. (2004)) could find support from the sunspot record.

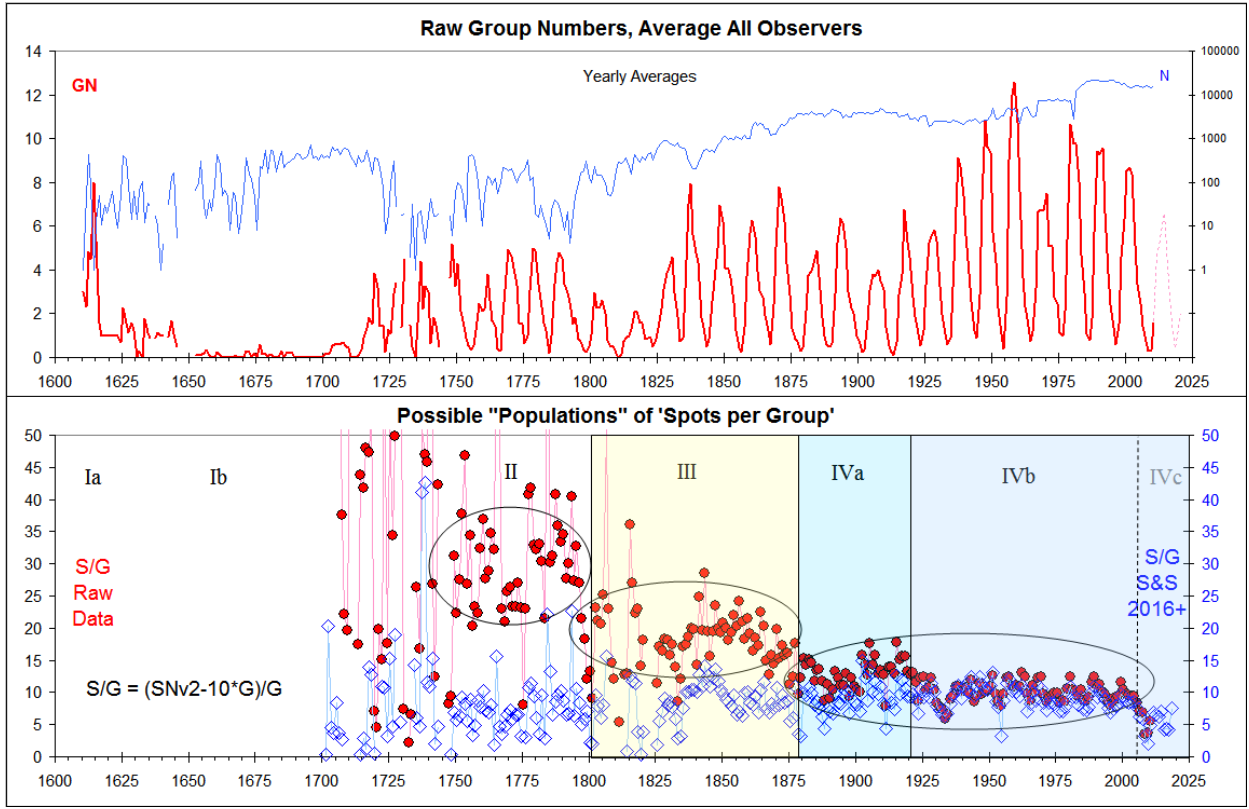

**Figure 1.** (Top) Yearly values of the number of groups averaged over all observers (red curve) without any normalization (i.e. 'raw' data; a few values off the top of the range are not plotted). The blue curve shows (on a logarithmic scale) the number of observations in each yearly average. (Bottom) Yearly averages of the number of spots per group (see text, section 3) calculated using the 'raw' average group numbers (red points) and the Svalgaard and Schatten (2016) group numbers (blue open diamond symbols). The ovals show various 'plateaus' suggesting several long-lasting 'regimes' of sunspot observing; also hinted at by different colored backgrounds.

2.1 How Many Spots in a Group?

With improving instruments and/or better observer understanding and interest, the number of spots (umbrae) observed in an active region (a 'group') increases, so the average number of spots in a group could be an indication of the 'quality' of the observation. On the other hand, and perhaps more important, at high solar activity it is difficult to decide which of the multitude of spots on the disk belong to a perceived group. The discoverer of the sunspot cycle, Heinrich Schwabe, reminded us that "Die schwierigste Aufgabe bei unsern Beobachtungen bleibt die Zählung der Gruppen" (Wolf, 1875). At high activity, groups blend together, so the number of groups reported by the observer tends to be too small (e.g. the count by the 18[th] century observer Staudach (Svalgaard, 2015), reported by Wolf and used by H&S). For those reasons, although the Number of Spots per Group cannot directly be used for calibration purposes, it is likely that if that number stays almost constant for a period of time, observing conditions (telescope, acuity, understanding, etc.) were also steady over that time, or, at least, if the number of spots per group should abruptly change to a different level we may have entered a different 'regime' of sunspot observing. For illustration, Figure 2 shows the ratio of spots to groups for observers Wolf and Wolfer using different telescopes and counting methods. You can see the combined effects of changing to a smaller telescope (78 mm → 37mm aperture) by Wolf, and of counting spots and groups differently by Wolfer (using the 83 mm Zürich telescope, almost equivalent with Wolf's 78 mm Bern telescope).

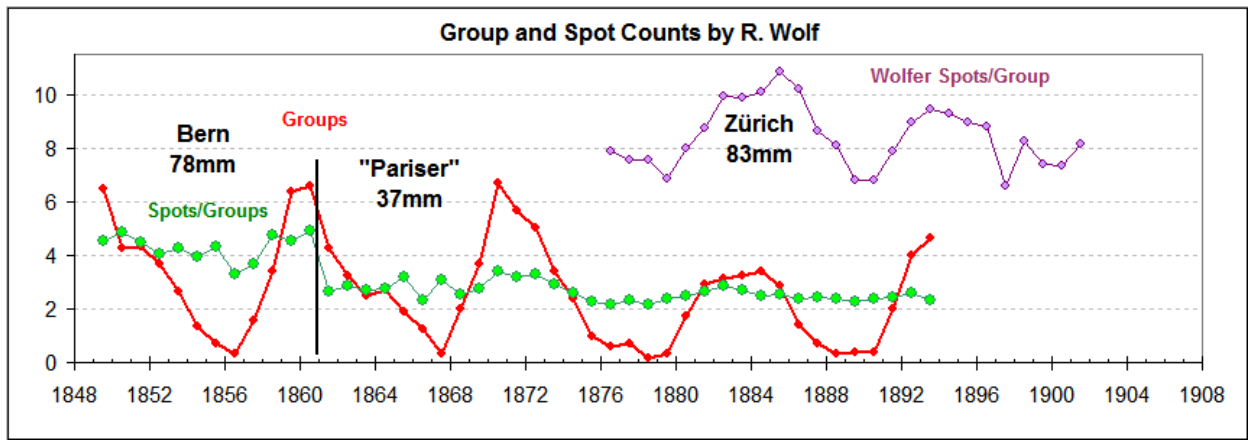

**Figure 2.** Yearly average number of spots per group for Wolf (green dots) and for Wolfer (smaller purple diamonds); data from the many *Mitteilungen* (https://www.ngzh.ch/) by the observers. Until 1861 Wolf used the 78 mm aperture telescope at Bern, but thereafter he used the much smaller (handheld) "Pariser" telescope with an aperture of only 37 mm. Wolf himself estimated that the sunspot number derived using the smaller telescope should be multiplied by 1.5 to be compatible with the larger telescope. As can be seen by comparison with the Group Number (red curve) there is a weak sunspot cycle dependence of the spot to group ratio.

## 3. Results and Discussion

When there are many observers, the 'combined' or average spot to group ratio is harder to evaluate as the number of *spots* seen by each observer is currently not readily available (which hopefully will change with the upcoming version 3 of the sunspot number). We shall here approximate that number, $S$, by using the Wolf formula $SN = k(10 \times G + S)$ on yearly values. For $SN$ we shall use version 2 of the sunspot number for which $k = 1$. For the group number, $G$, we can use, first, the yearly values of the raw group sunspot number, and, second, the yearly values of Svalgaard and Schatten (2016) group number, $GN$. We compute and plot in the bottom panel of Figure 1 the ratio $P = S/G = (SN - 10 \times G)/G$ for the two choices of $G$ ('raw' red; S&S blue). Ideally, the blue data points should all cluster around the same value (about or slightly lower than 10). They do not quite, but are close enough for our purpose here. We also gloss over the slight inconsistency caused by $G$ not being the same group number used in determining $SN$.

3.1 Populations of Solar Observations

On the other hand, the *P*-ratios for the raw averages of all observers cluster roughly in the three ovals shown in Figure 1 (apart from the earliest values with their large scatter). We shall refer to these different 'regimes' as different *Populations* of solar activity. Here we assume with Galton (1907) the usefulness of the "Wisdom of Crowds" to deal with observers of a common phenomenon, laboring under circumstances similar within populations, but different between populations (e.g. major improvement of instruments, such as the advent of achromatic lenses). We posit the existence of four major populations (I through IV; the first, during the Maunder Minimum (1645-1700), being totally conjectural at this point) with possible (speculative) sub-populations (a, b, … that are not the main focus of the present article). The *P*-ratio for populations II, III, and IV are approximately 30, 20, and 10, respectively (ignoring small differences between sub-populations). At first glance it seems strange that the ratio decreases with time, as instruments were supposed to get better with time. The reason is, of course, that the Zürich Compilers already strove to compensate for changing instruments and counting methods in their construction of the sunspot number, so *P* must be dominated by an *artificial* secular increase of the number of groups (being in the denominator), either reported by the observers or determined from their drawings of the spots on the disk. This conclusion hinges on the assumption that the sunspot number series (v2) is, at least, approximately 'correct', that is: a good indicator of solar 'activity', by which we today generally mean: manifestations of the 'solar magnetic field'.

3.2 Proxies for the Solar Magnetic Field

Fortunately, there are many manifestations of the solar magnetic field that can be used as 'proxies' for the activity. We take the view that a proxy, *Y*, of *X* is to be considered just another type of measurement of *X*, especially if the physical connection $Y = f(X)$ between *X* and its proxy *Y* is reasonably well-understood. Under this view, the sunspot number itself is just one of the proxies for (perhaps difficult to define 'true') solar activity; thus for us, "proxy" does not carry (as is often the case) any negative connotations. In the following we shall discuss several proxies that quantify interactions between solar activity and, particularly, the geomagnetic field (the latter, by the way, has been studied scientifically even longer than sunspots (Gilbert, 1600)), and show how they corroborate the modern sunspot series.

3.2.1 Solar Extreme Ultraviolet Radiation and the Diurnal Variation of the Geomagnetic Field

Graham (1724) discovered that the Declination, i.e. the angle between the horizontal component of the geomagnetic field and true north, varied through the day. Wolf (1852), and independently Gautier (1852), found it to vary with the number of sunspots. Thus was found a relationship between the diurnal variation of the geomagnetic field and the sunspots, "not only in average period, but also in deviations and irregularities", establishing a link between solar and terrestrial phenomena. Wolf soon found (Wolf, 1859) that there was a simple, linear relationship between the yearly average amplitude, *v*, of the diurnal variation of the Declination and his newly defined relative sunspot number, *R*: $v = a + bR$, allowing him to calculate the terrestrial response from his sunspot number. He marveled "Wer hätte noch vor wenigen Jahren an die Möglichkeit gedact, aus den Sonnenfleckenbeobachtungen ein terrestrisches Phänomen zu berechnen".

Stewart (1882) suggested that the diurnal variation was due to the magnetic effect of electric currents generated by daily movements across the Earth's magnetic field of an electrically conducting layer high in the atmosphere, in what we today call the ionosphere, but it would take another half century before the notion of conducting ionospheric layers was clearly understood and accepted: the E-layer conductivity starts to increase at sunrise, reaches a maximum near noon, and then wanes as the Sun sets, in accordance with the zenith angle of the Sun. Solar Extreme Ultraviolet (EUV) radiation with wavelength below 102.7 nm is the cause of the ionization in the E-layer, the resulting conductivity scaling with the square root of the overhead EUV flux (Yamazaki and Kosch, 2014; Svalgaard, 2016). The electric currents cause magnetic effects at the surface that are mainly felt at mid-latitudes in the East (*Y*) Component of the geomagnetic field, Figure 3, through a complex, but well-understood, chain of physical connections (see Figure 1 in Svalgaard (2016)).

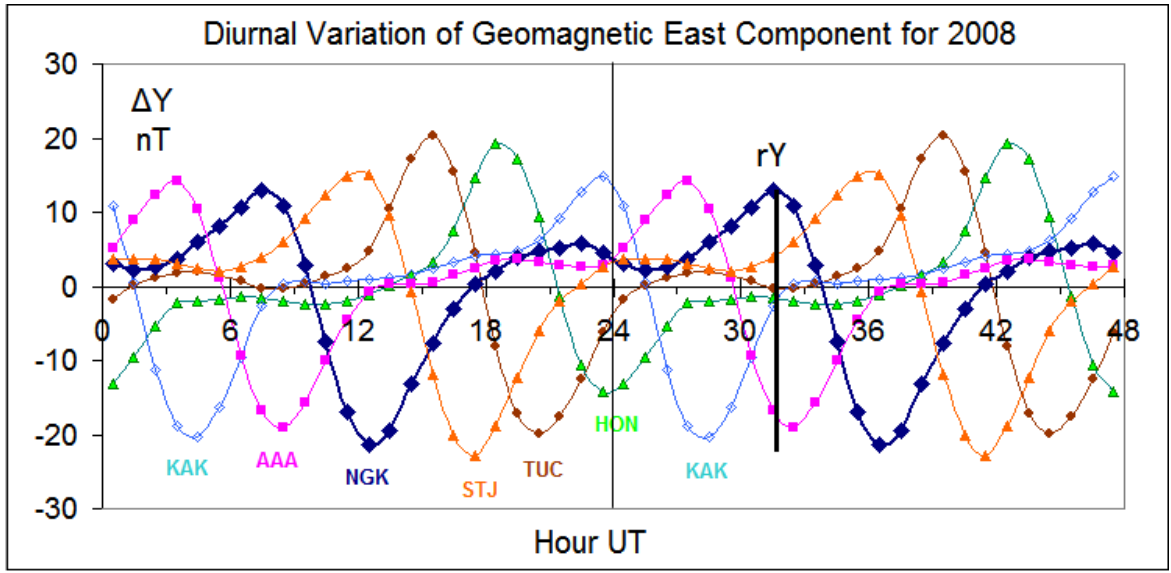

**Figure 3.** Average diurnal variation (over 48 UT hours in the low-activity year 2008) of the East Component, *Y*, at several geomagnetic observatories (http://www.wdc.bgs.ac.uk/catalog/master.html) spaced about 60 degrees apart in longitude, spanning the globe. The shape of the magnetic signature is remarkably stable; as we walk around the globe we note that the variation (deviation from the mean; in early parlance called the 'inequality') is almost the same from station to station, only differing very slightly in amplitude (*rY*, shown by the vertical bar), thus lending itself to a straightforward normalization (e.g. to Niemegk, NGK, as was done in Svalgaard (2016)).

The Diurnal Range, *rY*, of the variation can be determined with confidence from observatory data back to 1840 and estimated with reasonable accuracy about a century further back in time, Figure 4. Svalgaard (2016) used the data for more than 46 million hours from observatories all around the world to infer the EUV flux from the geomagnetic variations and found that normalized to measurements by spacecraft since 1996, the integrated EUV flux below 103 nm is well represented by $EUV = (rY/21.55)^2$ mW m$^{-2}$. Provided that we assume that the Sun has not changed just when we have the technology to look at it, we suggest that this relation holds generally and therefore affords a check on our reconstructions of solar activity (lest we should claim discovery of a new and unexpected solar variation).

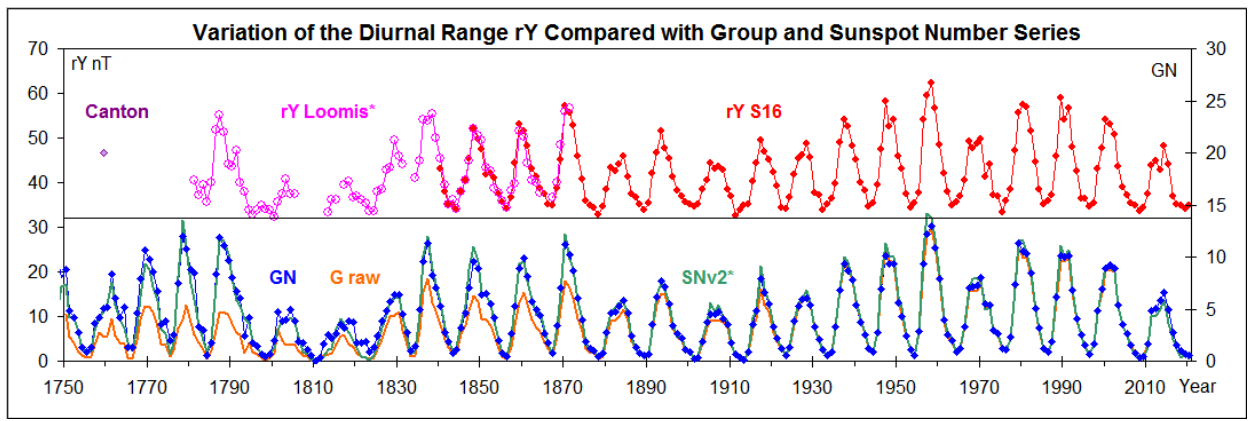

**Figure 4.** (Top) Yearly values of the diurnal range, *rY*, of variation of the East Component of the geomagnetic field as determined by Canton (1759), Loomis (1870; scaled to match Svalgaard (2016) during 1840-1870), and Svalgaard (2016). Where range in Declination was reported in arc minutes it has been converted to force units (nT) taken into account the secular change of the horizontal component. A slight (but not significant) increase in amplitude may be due to the change of the main magnetic field magnitude that impacts ionospheric conductivity (Cnossen and Matzka, 2016). (Bottom) The Group Number GN (Svalgaard and Schatten, 2016; blue symbols), the Sunspot Number SN version 2 scaled to GN (by dividing by 19; SILSO, 2020; green curve), and the 'raw' average group number of all observers (orange curve), as shown in Figure 1.

It is evident that the *rY* proxy for EUV agrees very well with the Svalgaard and Schatten (2016) reconstruction of the Sunspot Group Number as well as with the now official and widely agreed-upon version 2 of the Sunspot Number (which we really should call the "Wolf" number: "[O]n pourrait la nommer *Série de R. Wolf*, pour m'en assurer la propriété. On pourrait se moquer de cette prétention; mais puisqu'il existe des auteurs sans conscience on est forcé de défendre sa propriété", Wolf (1877)), and does not agree with the raw group data or the H&S series.

3.2.2 Heliomagnetic Field in Solar Wind Deduced from Geomagnetic IDV-index

Svalgaard et al. (2003) and Svalgaard and Cliver (2005, 2010) introduced a new geomagnetic index, the IDV-index, and showed that it was possible to infer with good accuracy the magnitude of the near-Earth heliospheric magnetic field all the way back in time to the invention of the magnetometer by Gauss and Weber and the burgeoning use at several observatories (≈1840: the "Magnetic Crusade"; Svalgaard, 2014). Although controversial at first (as breakthroughs seem to be), this is no longer the case (e.g. Lockwood and Owens, 2011; Owens et al., 2016a; Cliver and Herbst, 2018) and near-Earth heliospheric magnetic field values now form a well constrained dataset stretching back 175 years, Figure 5. The IDV-index measures the energy content of the Van Allen Belts around the Earth (the 'Ring Current') and has the useful property of depending directly on the strength of the solar wind magnetic field impinging on the geomagnetosphere.

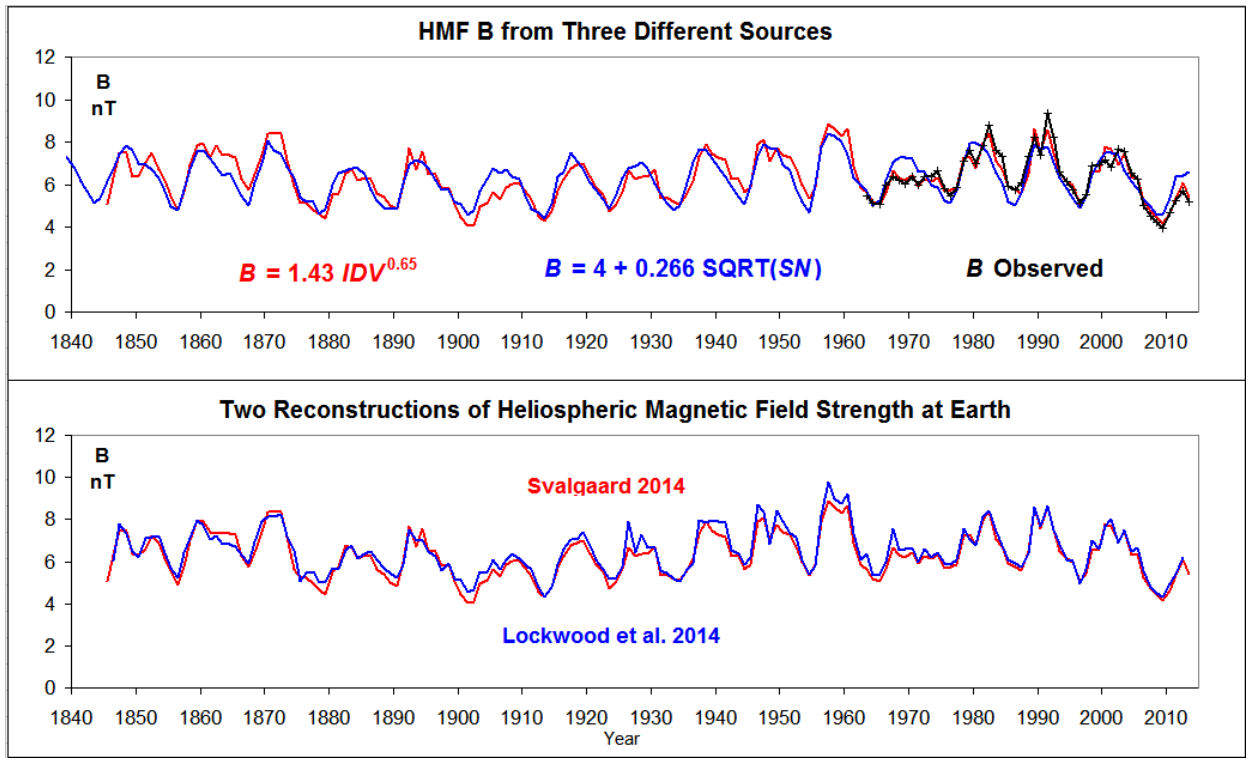

**Figure 5.** (Top) Heliospheric magnetic field, *B*, near the Earth inferred from the IDV-index (red curve), from the sunspot number (v2, blue curve), and observed by spacecraft (OMNI data https://omniweb.gsfc.nasa.gov/ow.html, black curve). (Bottom) The magnitude of the heliospheric magnetic field, *B*, inferred by Svalgaard (2014; red curve) and by Lockwood et al. (2014; blue curve, who now agree very well with *B* inferred by Svalgaard and co-workers). This is real progress.

The main sources of the low-latitude components of the Sun's large-scale magnetic field are large active regions. If these emerge at random longitudes, their net equatorial dipole moment will scale as the square root of their number. Thus their contribution to the average heliospheric magnetic field strength will tend to increase as the square root of the sunspot number (Wang and Sheeley, 2003) which fits well with the correlation shown in Figure 5.

3.2.3 Heliomagnetic Field in Solar Wind Deduced from Cosmic Ray-Created Radionuclide Data

It is also possible to reconstruct *B* using cosmogenic radionuclide data generated by cosmic rays, although somewhat less accurately than from geomagnetic and sunspot number variations arising from the fact that there are other influences on the cosmic ray flux at Earth and because the near-Earth heliomagnetic field *B* is a local measure of the heliosphere whereas the galactic sources of cosmic rays, having been generated in supernova explosions throughout the galaxy, influence the heliosphere as a whole. The cosmic ray record is further affected by terrestrial climate effects on the deposition in the reservoirs in which they are measured, and by geomagnetic field variability, variations in the local interstellar spectrum of cosmic rays, and high-energy solar energetic particle events. The data show that the cosmic ray intensity at Earth varies strongly throughout the solar cycle as a consequence of the varying structure (Svalgaard and Wilcox, 1976) and intensity of the heliospheric magnetic field (e.g. Potgieter, 2013). When hitting the atmosphere, the cosmic rays initiate cascades of nuclear reactions that lead to production of cosmogenic radionuclides subsequently sequestered in ice cores ($^{10}$Be) and tree rings ($^{14}$C), from which the heliomagnetic field strength can be inferred by suitable modeling. Figure 6 shows one such reconstruction by McCracken and Beer (2015).

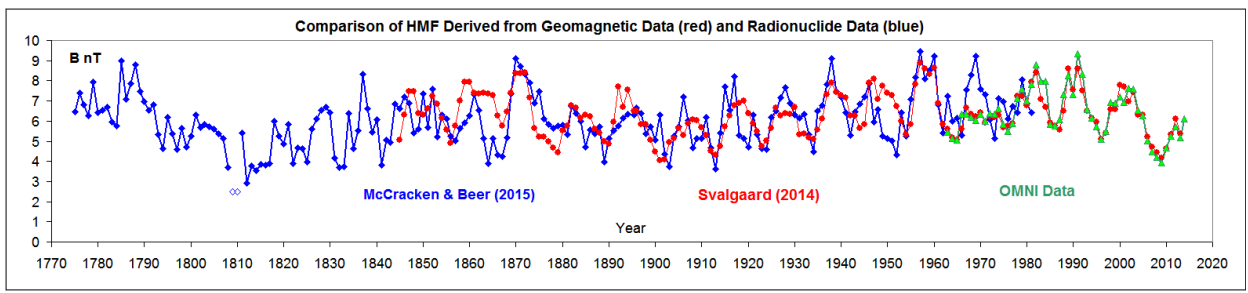

**Figure 6.** (Top) Yearly values of the heliospheric magnetic field, *B*, near the Earth inferred from the IDV-index (red symbols; Svalgaard, 2014), from the cosmic ray record (blue symbols; McCracken and Beer, 2015), and observed by spacecraft (OMNI data, green triangle symbols).

The process for converting $^{10}$Be concentrations in ice cores to *B* is more complex than with geomagnetic and sunspot data, and the uncertainties in *B* are thus larger. Nevertheless, there is good overall agreement between the cosmic ray-based *B* and the geomagnetic- and sunspot number-based series, especially when excising low values attributed to sporadic high-energy solar proton events. The reconstruction and agreement are discussed in detail by Owens et al. (2016b). We note, in particular, that the high activity in the 1780s and 1870s on par with recent 20$^{th}$ century activity (the "Modern [but not so *Grand*] Maximum") is well marked. The 'floor' in *B* (Svalgaard and Cliver, 2007) also looks supported and (now) sharper determined at *B* ≈ 4 nT.

3.2.4 The Waldmeier Effect

The director of the Zürich Observatory (1945-1979) Max Waldmeier (1978) reminded us that [for years 1849-1978] "there is a relationship between the rise time from minimum to maximum and the maximum smoothed monthly sunspot number. The times of the extrema can be determined without knowledge of the scale factor. Since this relationship also holds for the years from 1750 to 1848 we can be assured that the scale value of the relative sunspot number over the last more than 200 years has stayed constant or has only been subject to insignificant variations". So, the shape of the sunspot cycle curve, and thus the rise time from minimum to maximum, do not depend on the 'scale value' of the sunspot number. Determination of the rise time can therefore be used to check if the scale value has changed (Figure 7). Although Waldmeier today is credited with "the Waldmeier Effect" for the finding that large sunspot cycles have shorter rise times than do small cycles, this fact was known already to Wolf ("Greater activity on the Sun goes with shorter periods, and less with longer periods. I believe this law to be one of the most important relations among the Solar actions yet discovered." (Wolf, 1861)) and was seriously discussed around the turn of the 20th century (e.g. Wolfer (1902); and others) and taken as evidence for an 'eruption-type' sunspot cycle freed from "the shackles of unduly close adherence to harmonic analysis" (Milne, 1935), although the allure of 'oscillators' still rears it head today.

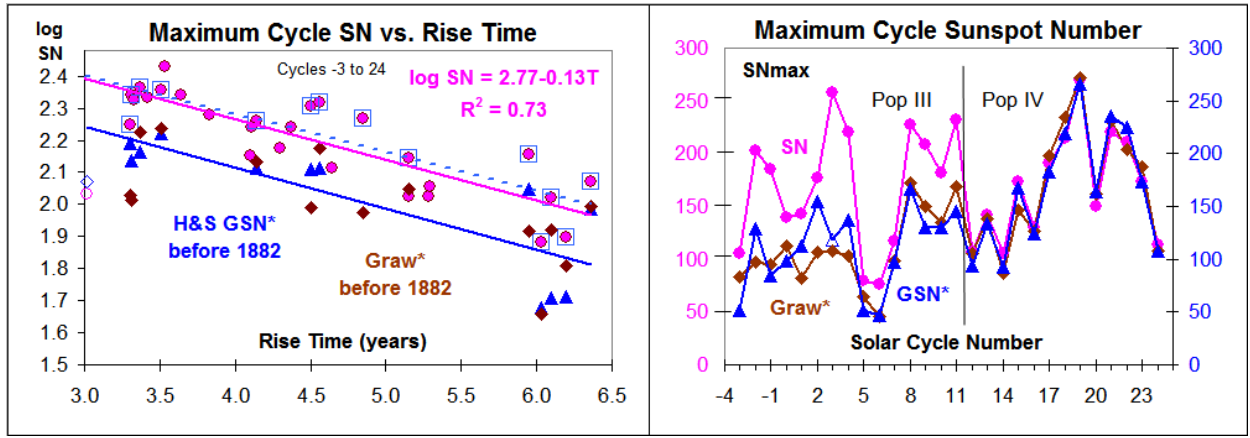

**Figure 7**. (Left) The logarithm of the maximum yearly sunspot number (v2) for solar cycles -3 to 24 (pink dots) with a pink linear trend line. The ones before 1882 (cycles -3 to 11) that we place in Population III are marked with a blue square with a dashed blue trend line. The H&S GSN can be scaled SILSO SN v2 based on data after 1882 (Population IV). Using the same scale factor (18.3) throughout we can put the logarithms of the scaled GSN* cycle maxima before 1882 on the plot (blue triangles) with a blue trend line. The difference in offsets (0.14) between the trend lines corresponds to a factor of 1.38 and is close to the disparity between Populations III and IV. The same exercise for the average group count ($G_{raw}$) of all observers (shown in Figure 1) with scale factor 21.65 yields the brown diamonds. (Right) The cycle maximum yearly sunspot numbers (pink dots) for cycles -3 to 24. Scaled cycle maximum group numbers for H&S (blue triangles) and for the average of all observers (brown diamonds) show the difference between Populations III and IV.

The Waldmeier Effect is also seen (as it should be) in other activity indices, such as sunspot areas and the ionospheric response to EUV (Svalgaard, 2020). There is no shortage of 'understanding' of the possible physical causes of the Waldmeier Effect (e.g. Kitiashvili and Kosovichev (2011); Karak and Choudhuri (2011); Russell et al. (2019)). In any event, the Waldmeier Effect is a firm observational constraint that any theory of the solar cycle must explain, and provides a solid underpinning for the calibration of the sunspot number.

If we define the 'growth rate', $g$, of a cycle as its maximum sunspot number, $SNmax$, divided by the rise time, $T$, the 'normal' Waldmeier Effect implies that $g = SNmax/T$ should also be larger for large cycles than for small cycles, and so it is: $SNmax = g \cdot T \sim T \exp(-T/2)$, Svalgaard (2020). This 2nd Waldmeier effect is actually statistically stronger than the 'normal' Waldmeier effect and is also found in the CaII emission of the sun and sun-like stars (Garg et al., 2019).

## 4. Comparisons with H&S

Cliver and Ling (2016) tried to reproduce the determination of the $k$-values determined by Hoyt and Schatten (1998) for observers before 1883 and failed because the procedure was not described in enough detail for a precise replication; in particular, it is not known which secondary observers were used in calculating the $k$-factors. On the other hand, H&S in their construction of the Group Sunspot Number did not use daisy-chaining (i.e. secondary observers) for data after 1883 because they had the Royal Greenwich Observatory (RGO) group counts as a continuous (and at the time believed to be good) reference with which to make direct comparisons. During the early years of the RGO data, the group counts were drifting (Cliver and Ling, 2016), but for the years after about 1900 when the RGO drift seems to have stopped or, at least abated, the H&S Group Sunspot Numbers agree extremely well with the Svalgaard and Schatten (2016) Group Numbers (with a scale factor of 13.6; Figure 8), and incidentally also with the various Lockwood and Usoskin reconstructions ("$R_{UEA}$ is the same as $R_G$ after 1900" Usoskin et al (2016)), and even with the (suitably scaled) revised sunspot number version 2.

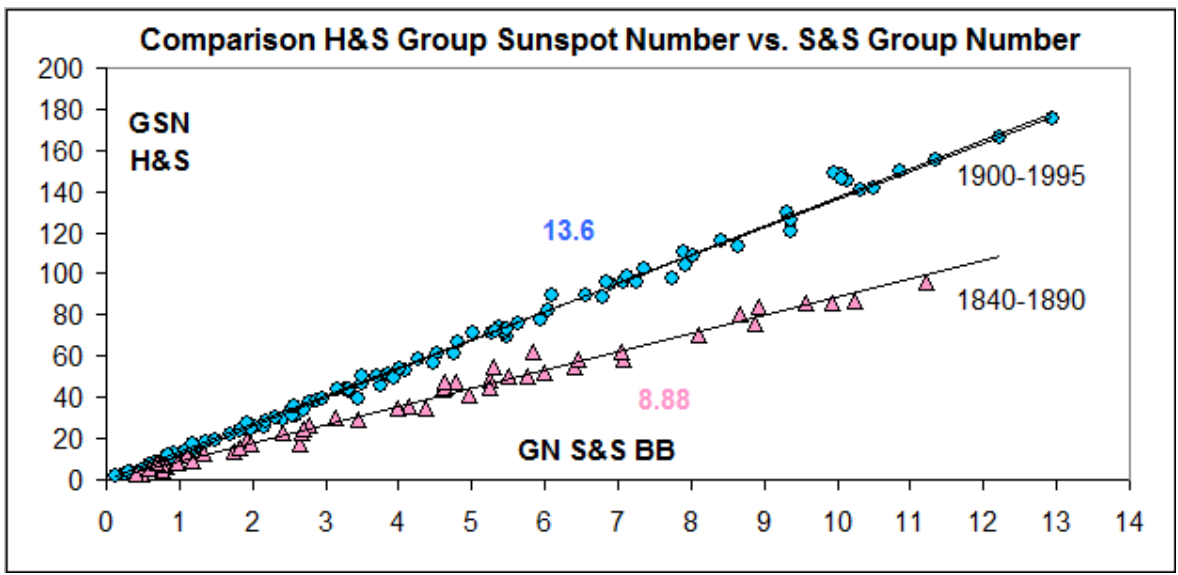

**Figure 8**. Annual averages of the Hoyt and Schatten (1998) Group Sunspot Number (GSN; often called $R_G$) compared to the Svalgaard and Schatten (2016) Group Number (GN). For the data since 1900 (light-blue dots) there is a constant proportionality factor of 13.6 between the two series. For earlier years, the drift of the RGO counts combined with daisy-chaining the too-low values back in time lowers the factor to 8.88 (pink triangles).

For the years 1840-1890 there is also a strong linear relationship, but with a smaller slope because the drift of RGO has been daisy-chained to all earlier years (Lockwood et al. (2016): "Because calibrations were daisy-chained by Hoyt and Schatten (1998), such an error would influence all earlier values of $R_G$", which indeed it did). The factor to 'upgrade' the early part of the series to the 'RGO-drift-free' part is $13.6/8.88 = 1.53$. Figure 9 shows the result of 'undoing' the damage caused by the RGO drift. H&S did not discover the RGO drift because their $k$-factor for Wolf to Wolfer (inexplicably) was set as low as 1.02, i.e. Wolf and Wolfer were assumed to see essentially the same number of groups relative to RGO and to each other, in spite of Wolf himself using a factor of 1.5 (albeit for the relative sunspot number of which the group number makes up about half). It is possible that this was due to not noticing that Wolf changed his instrument to a smaller telescope (c.f. Figure 2) when he moved to Zürich (as the larger 'norm-telescope' had not been delivered yet).

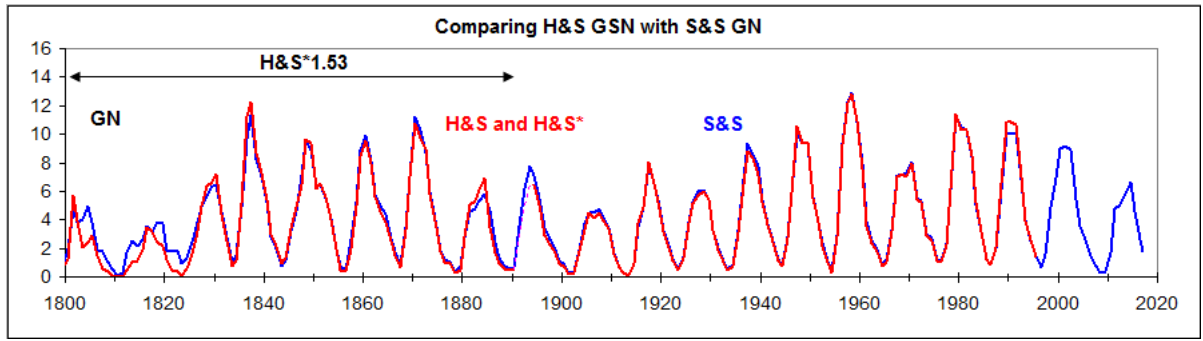

**Figure 9**. Annual averages of the Hoyt & Schatten (H&S) Group Sunspot Number divided by 13.6 (red curve) since 1900 compared to the daisy-chain free Svalgaard & Schatten (2016) Group Number (S&S GN, blue curve). For the years 1800-1890, the H&S values were then scaled up by 13.6/8.88=1.53. This brings H&S into agreement with S&S, effectively undoing the damage caused by the single daisy-chain step at the transition of H&S from the 19$^{th}$ to the 20$^{th}$ century.

If it were not for the mistake of using a *k*-factor of 1.02 for Wolf instead of the actual 1.66 we conclude that H&S's GSN was actually pretty good back to 1800 AD, basically agreeing (after correction) with the Svalgaard and Schatten (2016) Group Number and the SILSO Sunspot Number version 2 (on the group number scale). How did it fare before Population III, i.e. before 1800 AD?

4.1 Calibration with 'Antique' Telescopes Before 1800 AD?

Our knowledge of solar activity during Population II in the 18$^{th}$ century centers on the observations by the amateur astronomer Johann Casper Staudauch who made more than 1100 drawings of the spotted solar disk (Svalgaard, 2017). Achromatic telescopes were manufactured in the late 1750s. With such an (expensive) telescope, however, the distinction between umbra and penumbra should have been clear, and the Wilson effect (elongated spots near the limb) should have been visible. Both were not drawn by Staudach (using projection onto a sheet of paper). Arlt (2008; Arlt and Vaquero, 2020), who currently curates the Staudauch drawings, suggests that Staudach missed all the tiny A and B spot groups (according to the Waldmeier classification). Such groups make up 30-50% of all groups seen today. Haase (1869) also reviewed the Staudach material and reports that a 4-foot telescope was used, but that it was not of particular good quality and especially seemed not to have been achromatic, because he quotes Staudach himself remarking on his observation of the Venus transit in 1761 that "for the size and color of the planet there was no sharp edge, instead it faded from the same black-brown color as the inner core to a still dark brown light red, changing into light blue, then into the high green and then to yellow".

So we may assume that the telescope suffered from spherical and chromatic aberration. We can build replicas with the same optical flaws as telescopes available and affordable to amateurs in the 18$^{th}$ century. On Jan. 16, 2016 we started observations of sunspots with such replicas. Three observers (expert members of "The Antique Telescope Society", http://webari.com/oldscope/)

have made drawings of the solar disk by projecting the sun onto a sheet of paper. We count the number of individual spots as well as the number of groups they form. Comparing our counts with what modern observers report for the same days we find that the sunspot number calculated from the count by modern observers is three times larger as what our intrepid observers see (Figure 10), and that the number of groups is 2.5 times as large.

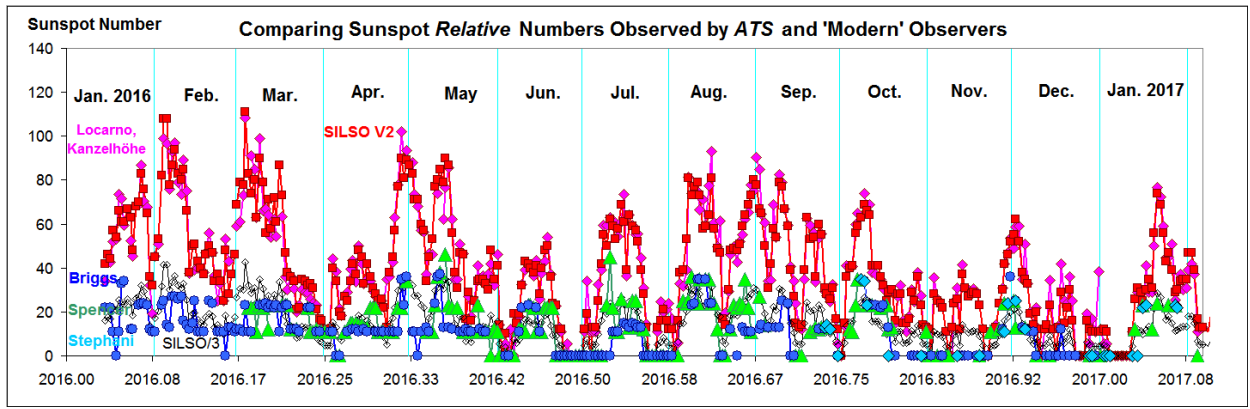

**Figure 10**. Daily observations of sunspots made with replicas of 18[th] century telescopes by members of the Antique Telescope Society (J. Briggs, blue dots; K. Spencer, green triangles; W. Stephani, light blue diamonds) compared with modern sunspot numbers (SILSO v2, red squares; average of Locarno and Kanzelhöhe observatories, pink diamonds). Dividing SILSO data by three brings the official sunspot number down to match the replica values (thin black line with open diamonds).

This suggests that we can calibrate the 18[th] century observations in terms of the modern level of solar activity by using the above factors. SILSO v2 SN divided by 3 (thin black curve on Figure 10) is a reasonable match to the sunspot number calculated from Staudach's drawings (Svalgaard, 2017) thus roughly validating the revised SILSO values and not compatible with the low values of the H&S reconstruction or with reconstructions that resemble H&S's. When Solar Cycle 25 picks up, we hope to restart the experiment and refine the calibration.

## 5. Solar Cycle 11: The Test Case

We have presented numerous, detailed arguments in favor of existence of several distinct populations of sunspot observations over time, but recognize that their number may exceed the Hrair-limit for many researchers (often referred to as our 'users') for whom mind-numbing minutia about data sets are on the periphery of their sphere of interest. As Cliver (2017) pointed out, we now have basically two classes of reconstructions: 1: A set of series that closely resemble the original Hoyt and Schatten reconstruction (which even Schatten knows is wrong) and 2: A set of series that closely resemble the 'official' Sunspot series (both versions; v2 is essentially just v1 divided by 0.6) and the closely agreeing Svalgaard and Schatten (2016) series. The main difference is (as already pointed out by H&S) a discontinuity around 1882 with up to 40% discrepancy between the two classes. The two classes largely agree going back in time until we come to Solar Cycle 11, peaking in 1870; the disagreement then persisting for all earlier cycles. Einstein famously said "No amount of experimentation can ever prove me right; a single experiment can prove me wrong"; so if a reconstruction does not get Cycle 11 right, it is wrong and should be discarded. Reconstruction of Cycle 11 can thus serve as that single experiment that every reconstruction must pass.

Chatzistergos et al. (2017), daisy-chaining too many, too short "backbones", advocate their reconstruction of the group number as "robust", leading Petrovay (2020) to hesitantly suggest that it "seems to be the most recommendable version for further analysis". Willamo et al. (2017)

optimistically see their Active-Day-Fraction based reconstruction "forming a basis for new studies of the solar variability and solar dynamo for the last 250 yr", although further testing paints a bit less rosy picture of their effort (Willamo et al., 2018). Dudok de Wit et al. (2019) present "a new approach that bypasses the need for intercalibration and in addition avoids the artificial introduction of backbone observers", although admitting that more testing is needed. These 'modern' reconstructions (which seem to be holding up completion of the ISSI Team 417 task) all belong to Cliver's class 1 that resemble the original H&S reconstruction. Figure 11 shows how they fare in representing Cycle 11.

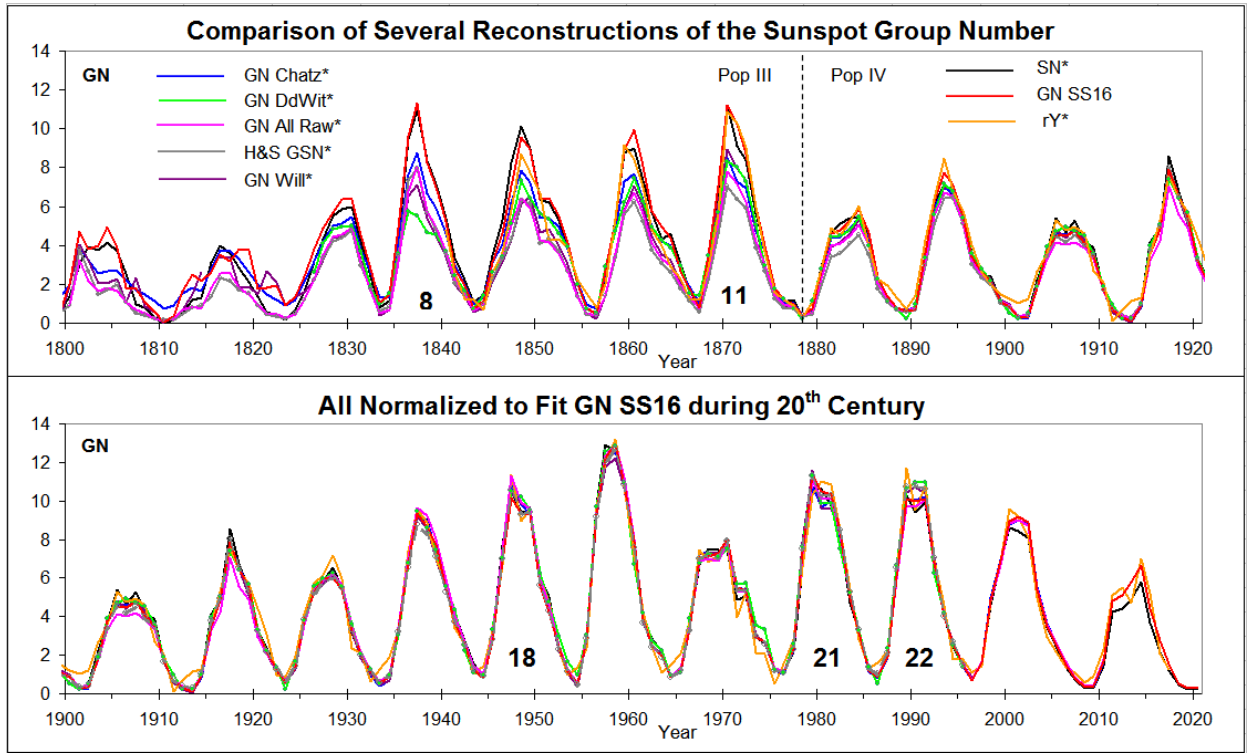

**Figure 11**. Comparison of reconstructions of the Group Number covering the transition between Population III and Population IV with special emphasis on Solar Cycle 11. Since all reconstructions agree during the 20$^{th}$ century we can for ease of comparison normalize them all (regression coefficients of determination are very high, varying between 0.96 and 0.99+) to the Svalgaard and Schatten (2016) reconstruction (red curve). The lower panel shows the result of the normalization since 1900 AD. All curves in that panel overlap so closely that it is difficult to see the individual reconstructions (shown with the color coding found in the upper panel; as green is problematic for colorblind readers, the green curve is marked with small dots for easier recognition).

In the upper panel, we can clearly see the distinction between Cliver's two classes. Class 2 is represented by the, mutually closely agreeing, Svalgaard and Schatten (2016) reconstruction (GN, red curve), the scaled sunspot number (SN version 2, black curve), and the scaled diurnal variation of the geomagnetic East Component (rY, orange curve) series. Class 1 is represented by the, approximately mutually agreeing (albeit with some scatter), Chatzistergos et al. (2017, blue), Dudok de Wit et al. (2019, green; based on the primary observer Schwabe who had a nonlinear response), Willamo et al. (2017, purple), original H&S GSN (gray), and raw average of all observers (pink, refer to Figure 1) series. It is evident that the class 1 series do not match Solar Cycle 11, and thus fail the critical test that any reconstruction must pass.

5.1 A Closer Look at Solar Cycle 11.

Because Solar Cycle 11 is so important it pays to take a closer look. Svalgaard (2019) constructed an improved backbone for Wolfer with monthly resolution, thus taking advantage of the much larger number of degrees of freedom available for the regression to determine the calibration coefficients. Figure 12 shows the coverage chart for the 30 observers who overlap directly with Wolfer and 'approaching Cycle 11 from above'.

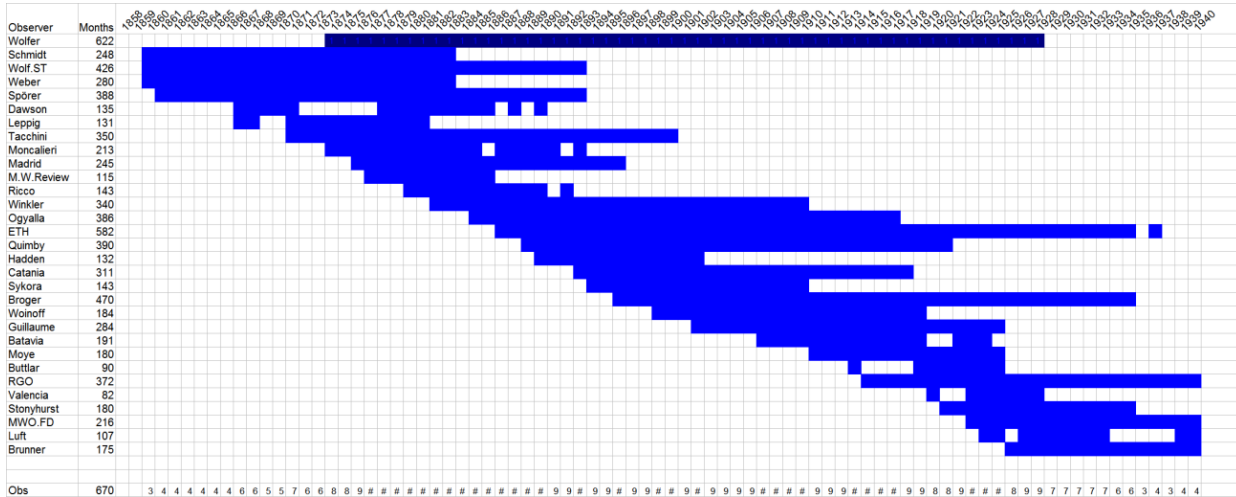

**Figure 12**. Wolfer observations began in 1874 and ended in 1928 comprising 622 months of data (where a month was counted if there were at least five observations). For each of the 30 observers a bar is drawn from the first year with a month with data to the last year with a month of data, omitting years with no observations.

For each observer and for each month with at least five observations, we calculate the monthly means and regress Wolfer's data against each observer's. The correlations are invariably very nearly linear with offsets that are not, or barely, statistically significant. Ignoring the offsets, we get for each observer a *k*-value as the slope of the regression line going through the origin, so can normalize the data for each observer to the Wolfer scale by simple multiplication by the appropriate *k*-factor. Figure 13 shows two example plots; for more see Svalgaard (2019).

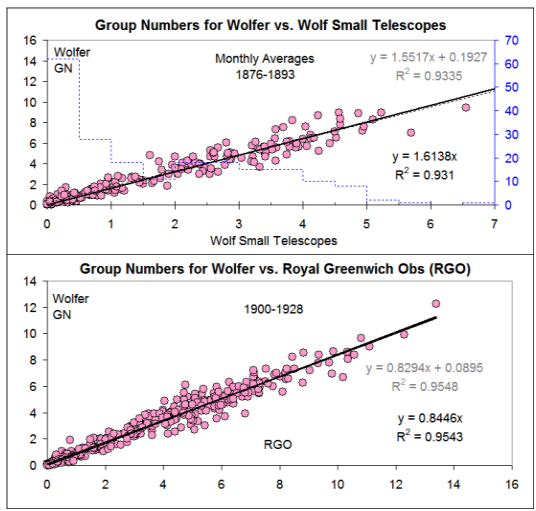

**Figure 13**. (Top) The monthly average group numbers for Wolfer against the corresponding values for Wolf using his small telescopes. Two regression lines are shown; one with and one without an offset going through the origin. They just happen to fall on top of each other because the offset is negligible. (Bottom) The same, but for the Royal Greenwich Observatory (RGO) for times after the drift has abated.

We can now reconstruct the composite group number backbone on the Wolfer scale for the years 1858 through 1940, Figure 14. As all comparisons of observers with Wolfer are *direct* without using intermediate observers there is no daisy-chaining.

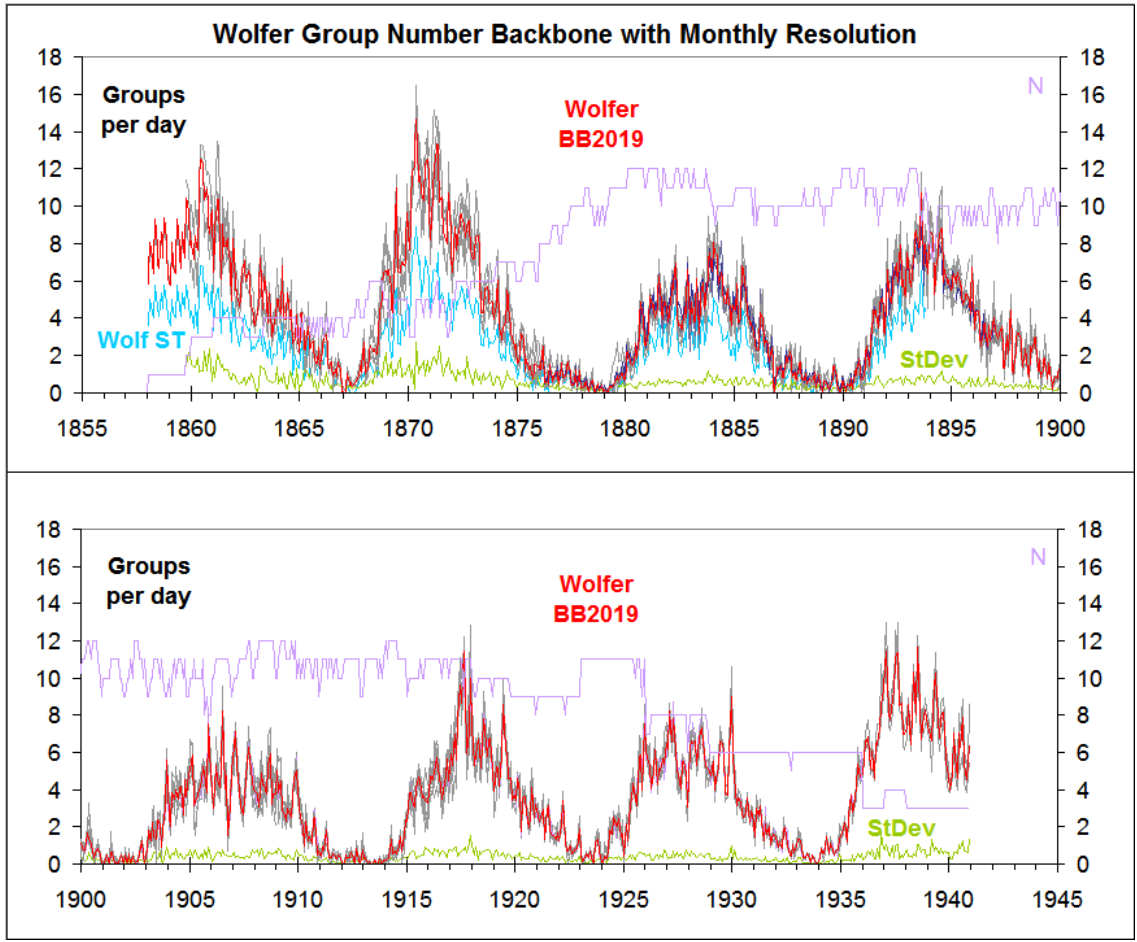

**Figure 14**. New Wolfer Backbone showing the average number of groups on the disk per day with monthly resolution. The individual observers' scaled data are plotted with gray curves while the average is plotted in red. The standard deviation is plotted in light green at the bottom of the figure. The number of observers for each month is shown by the light purple step line. Wolf's observations with his small handheld telescopes are shown by the light blue curve.

We can compute yearly values from the monthly values and compare with the Svalgaard and Schatten (2016) Group Number Backbone, Figure 15. The agreement is excellent, so Cycle 11 is well in hand on the scale of Population IV.

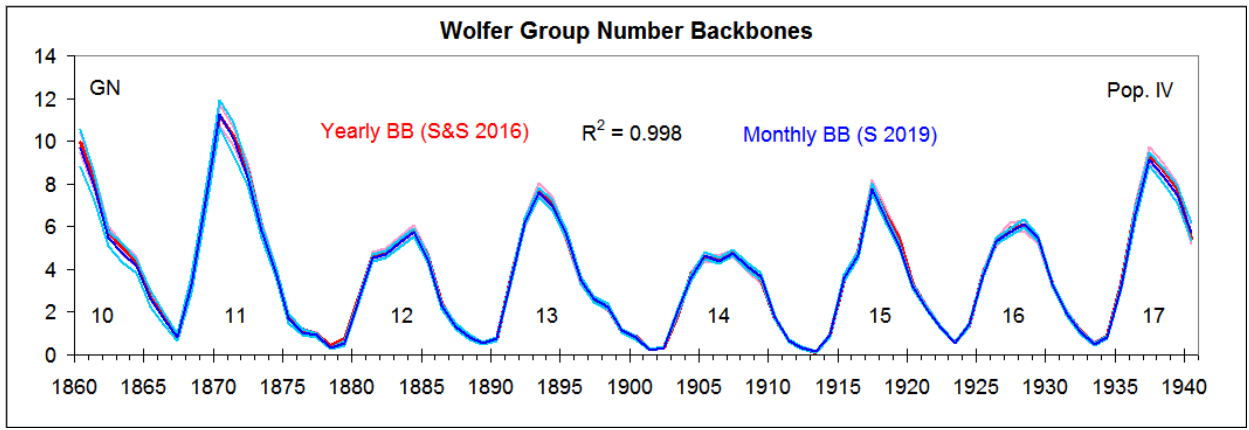

**Figure 15**. Comparing the monthly Wolfer Backbone (blue) with the yearly Wolfer Backbone (red). One sigma error bands (with lighter color) surround the curves.

# 6. On the Wisdom of Crowds

We have already remarked on the uncanny fact that the raw average of all observers bears a strong resemblance to the painstakingly normalized reconstructions. This is especially true when the different populations are taken into account, i.e. when the backbones stay within populations. We illustrate this in Figure 16 covering three major backbones since 1800 AD.

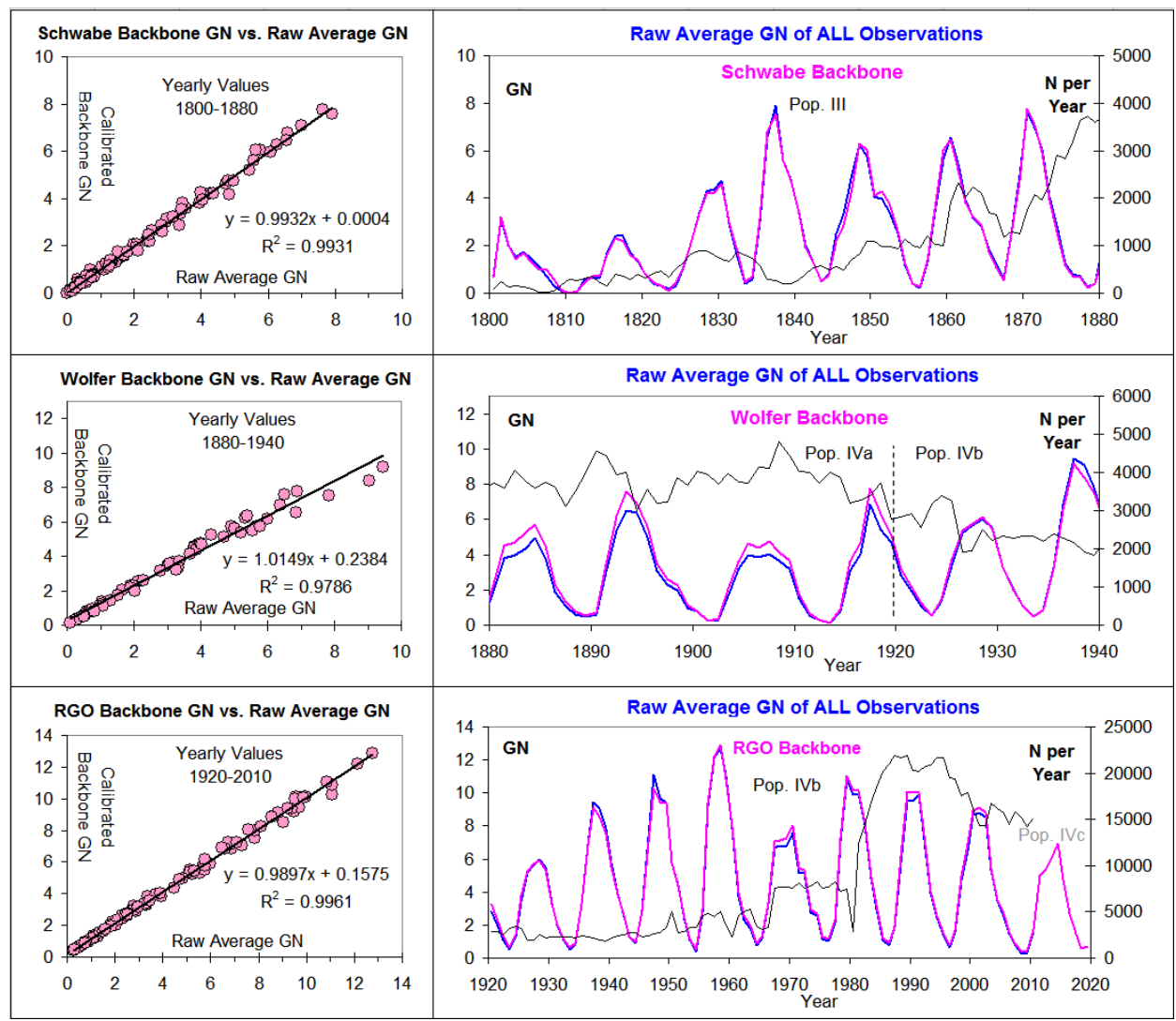

**Figure 16**. Comparing the part of the yearly (top) Schwabe Backbone that falls within Population III to the raw average of all observers, (middle) Wolfer Backbone that falls within Population IV, and (bottom) RGO Backbone that falls within Population IV. The left-hand panels show the calibrated backbone group numbers regressed against the raw average group numbers. The correlations are all linear (and show close proportionality due to negligible offsets) with slopes within 1% of unity. The right-hand panels compare the normalized backbones (pink) to the raw averages of all observers (blue). The thin black curves show the number of observations in each year (on the right-hand scale), generally numbering in the thousands.

It appears that the Wisdom of Crowds (Galton, 1907; Aristotle, 350 BCE, Politics, III:xi) works amazingly well. This may mean that we can dispense with the normalization altogether (although adjacent, overlapping backbones (of which we only have two, back to 1800 AD, one on each side of the Wolfer backbone) still have to be stitched together by pair-wise comparison without any intermediaries, and thus not even with daisy-chaining). If so, it seems (perhaps with 'tongue in cheek') that we may have a nice non-parametric, non-overlapping, no k-value-regression, no selection effect, no tied-ranking, no daisy-chaining, no ADF- or PDF-based, no-whatever method for constructing a backbone segment including estimating its time-varying error bars (from the spread of the observations). Or it may mean that no matter what we do, the results will approach the raw averages (when thousands of independent observations are involved, Galton (1907)) and that it therefore is not surprising that they do. Only by realizing that there are several populations and that we need independent proxies (like geomagnetic *rY*, HMF *B*, observations with telescope replicas, and cosmic ray radionuclides) to overcome the artificial discontinuities between populations can we make progress towards that elusive goal: a unified and vetted set of solar activity indices that can be universally accepted and used, instead of being a moving target and a free parameter.

## 7. Conclusion

We have covered a lot of ground to support the simple assertion that the (already basically agreeing) revisions of the sunspot relative number (should better be known as the Wolf number) and of the sunspot group number series put forward in the epoch-making report by Clette et al. (2014) were significant steps forward in curing the ills of the disparate solar activity datasets then in use. Fortunately, the revisions spawned extensive discussion and research into the basis, data, and construction of the indices; something the old series (basically accepted on faith or, at times, expediency) sorely lacked. But, such a period of 'soul-searching' and, especially, dissent sows confusion and undermines the usefulness of the very concept of quantitative measures of solar activity and should eventually come to an end. Luckily, the findings now at hand and reported here suggest that the time has come for the long-awaited recognition of our resolution of the conundrum holding back or delaying progress in a field so important for our understanding of the Sun, with attendant societal consequences for a space-faring civilization.

### Acknowledgments

The author acknowledges the use of data from the Historical Archive of Sunspot Observations at the Library at Centro Universitario de Mérida, Spain, of data from WDC-SILSO, Royal Observatory of Belgium, Brussels, and of the OMNI data from NASA. The author thanks Phil Scherrer at Stanford University for continued support and declares to have no financial conflicts of interest.